# Multiphase Hydrogen Storage in Nanocontainers

Suboohi Shervani[1], Puspal Mukherjee[2], Anshul Gupta[1], Gargi Mishra[3], Kavya Illath[4], T. G. Ajithkumar[4], Sri Sivakumar[3], Pratik Sen[2], Kantesh Balani[1] and Anandh Subramaniam[1*]

[1]*Department of Materials Science and Engineering, Indian Institute of Technology, Kanpur-208016, India.*
[2]*Department of Chemistry, Indian Institute of Technology, Kanpur- 208016, India.*
[3] *Department of Chemical Engineering, Indian Institute of Technology, Kanpur -208016, India.*
[4]*Central NMR Facility, CSIR-National Chemical Laboratory, Pune-411 008, India*
[*]anandh@iitk.ac.in *(author for correspondence)*

**Abstract**

Hydrogen can be stored in containers or in materials. In materials it can exist in molecular or atomic forms. The atomic form can further exist as multiple phases. Molecular hydrogen can be adsorbed on the surface or can be present inside the material. In the current work, we demonstrate for the first time, a methodology to store hydrogen in nearly all conceivable forms. At the heart of this strategy is the novel concept of storage of molecular hydrogen in metal (Pd) nanocontainers. The existence of multiple phases of hydrogen is confirmed by x-ray diffraction, Raman spectroscopy, nuclear magnetic resonance and pressure-composition isotherm measurements. About 18% of the hydrogen stored is in the molecular form and interestingly this is achieved at room temperature and 1 atm pressure. Enhancement in storage capacity as compared to nanocrystals of the same mass is observed (36 % increase at 1 atm & 25°C).

Hydrogen holds considerable promise as a fuel of the future[1]. In the utilization of hydrogen as a fuel, its storage and release are important steps[2]. Hydrogen can be stored in multiple forms: in molecular form (as free molecules in containers or adsorbed on the surface of materials) or in atomic form in a host lattice. Storage of hydrogen in materials has attracted considerable attention in the past few decades and multiple classes of materials have been investigated in this regard. These include metals and alloys, zeolites, metal organic frameworks (MOFs), clatherates and carbonaceous materials[3]. Typically each of these materials store hydrogen in one or two states/phases; e.g., in metals it is stored as one or two solid solutions or a solid solution and a hydride (compound) and in MOFs it is stored in entrapped molecular form (inside the material). A specific class of these materials can offer us certain benefits in terms of hydrogen storage, but correspondingly may suffer from some other lacunae. Magnesium has good storage capacity, but high temperatures are required for storage[4]. Carbonaceous materials can also offer good storage, but usually at low temperatures (liquid nitrogen)[5]. Hybrids are emerging as an important class of materials, wherein the synergistic benefits of two or more components can lead to better hydrogen storage properties[6-8].

Nanoscale materials are natural choice as hydrogen storage materials, due to fast kinetics of absorption and desorption and enhanced surface and sub-surface storage of hydrogen. Nanostructured materials developed for hydrogen storage include: carbonaceous materials[5] (e.g. carbon nanotubes, carbon nanofibers & graphene), nanoparticles (e.g. MgO[9], Pd[10]), nanostructured bulk materials (e.g. ZK60 Mg Alloys[11]) and nanohybrids (e.g. Pd@MOF[6], Pd@Pt[12]). Molecular cages have also been used to store hydrogen[13]. Hollow nanospheres have been investigated in diverse contexts[14-16]. In in-press manuscript the importance of Pd hollow spheres for hydrogen storage is highlighted using molecular dynamics[17].

In the current work we demonstrate a novel strategy to store hydrogen in nearly all conceivable forms: (i) in a container, (ii) atomic form as a solid solution, (iii) in molecular form on the surface (adsorbed). This multi-pronged strategy can be exploited in the future for designing materials with highly enhanced hydrogen storage capacity; with the benefit of fast kinetics of absorption and desorption.

To achieve the abovementioned aim, Pd hollow spheres were synthesized by first producing silica nanospheres (solid core with mesoporous shell, SCMS) by a modified Stober's method[18], followed by formation of Pd shell using chemical reduction (sample-A1). Sample-A1 was processed via two routes: (i) evacuation at 80°C for 12 hours, followed by annealing at 200°C to seal the pores (sample-A2), (ii) the inner silica core was etched out by sodium hydroxide to give a porous Pd shell (sample-A3). The porous Pd hollow nanospheres (PdHS, sample-A3) was evacuated at 80°C for 12 hours, followed by annealing at 200°C to seal the pores and to obtain a non-porous PdHS (sample-A4). Figure 1 shows TEM micrographs obtained from sample-A1 and sample-A3. Inset to Figure 1b shows the selected area diffraction pattern (SADP) obtained from sample-A3. The porous nature of the shell is to be noted and the rings indicate the nano-polycrystalline nature of the Pd shell.

Two strategies are used to 'entrap' molecular hydrogen in the nanocontainers. Hydrogenation of the sample-A3 by pressurization with hydrogen at 20 bar at 80°C for 6 hrs, followed by heating at 200°C for 2 hrs (leading to the sample designated as A3b). This treatment is expected to fill hydrogen in the PdHS via the pores followed by the sealing of the pores. Sample-A4b was obtained from sample-A4 by first annealing at 200°C for 2 hrs at $10^{-5}$ Torr, followed by pressurizing in hydrogen atmosphere at 20 bar pressure for 6 hrs (200°C).

Pressure-composition isotherms (PCI) were obtained using a Sieverts apparatus, wherein the pressure of hydrogen was increased from $10^{-3}$ bar to 1 bar (at 25°C) and the amount of hydrogen absorbed by the material was measured. Two kinds of samples were tested in the PCI apparatus (Figure 2): (a) with $SiO_2$ core (sample-A1) and (b) non-porous PdHS (obtained from sample-A4 by annealing at 300°C for 6 hrs at $10^{-5}$ Torr ). The PCI curves were obtained after running two cycles of absorption and desorption (akin to the practice for Pd nanoparticles[19]). Kinetics of absorption (wt.% hydrogen absorbed as a function of time at 25°C and 1 bar pressure) is shown as inset to Figure 2. The weight percent in these figures is calculated as: [(wt. of hydrogen absorbed)×100]/[(weight of (Pd+Hydrogen)] and the higher weight percent of hydrogen stored in the non-porous PdHS sample (of 0.61%) is to be noted. The mass of Pd is identical in both the samples (hollow versus filled) and hence the additional hydrogen stored (of 0.11 wt.%) must be inside the nanocontainer. This value is higher than that obtained for Pd nanoparticles (of 0.45 wt.%, ~7 nm[10]).

Figure 3 shows the XRD results from sample A3b hydrogenated to: (a) 1 atm and (b) 0.02 atm. The formation of a single phase β interstitial solid solution of hydrogen in Pd crystal is seen in Figure 3a, while a mixture of α & β solid solution phases is seen in (b). In nanoscale Pd, it is reported that in the α-phase the hydrogen is present in the octahedral voids, while in the β-phase 30% of the hydrogen is present in the tetrahedral voids[20]. Thus we observe that hydrogen in atomic form can be stored in more than one phase, with specific differences in the type of void involved.

Whole Profile Fitting (WPF) was carried out using JADE (MDI Corp) to determine the lattice parameters and phase fractions of the two phases (calculations in Supplementary Information, Section 2.1). The lattice parameters of the α and β phases were determined to be 3.893 Å and 4.023 Å, respectively and the corresponding phase fractions are 77 wt.% and 23 wt.% for the XRD pattern shown in Figure 3(b). The lattice parameter of the α phase is slightly expanded as compared to pure PdHS (with $a_{PdHS}$ = 3.890Å). The composition of the α phase (~2 at.%H, $PdH_{0.02}$) and β phase (~36 at.%H, $PdH_{0.57}$) were calculated using the relation $a^3 = 10.5x + a_0^3$, where $a_0$ represents lattice parameter of pure PdHS and x represents amount of hydrogen (at.%)[21]. The amount of hydrogen obtained in the α phase is lower than that in nanocrystals[22] (4-6 nm, ~$PdH_{0.05}$) but close to that of bulk[23] ($PdH_{0.60}$), whereas for β phase it is lower than that obtained in bulk but higher than that in nanocrystals[22] ($PdH_{0.46}$).

Raman spectroscopy results from the non-porous PdHS (samples-A3b & A4b) are shown in Figure 4. To understand the origin of the peaks in the Raman spectra, simulations were carried out using Gaussian software. The Raman signature of molecular hydrogen was analyzed using two kinds of models: (i) free molecular hydrogen, (ii) hydrogen adsorbed on Pd surface ((111), (110) & (100)). The details of the computation along with the models used can be found in the supplementary material (Section 1.6). The results of the computations are overlaid on Figure 4, which shows the existence of molecular hydrogen in three forms: (i) free, (ii) adsorbed on Pd surface and (iii) bilayer adsorbate. It is interesting to note that in spite of the complications arising from the myriad of possibilities with respect to the adsorption of hydrogen[9] (different surfaces, surface ledges/kinks, surface curvature, etc.) sharp Raman peaks are obtained. A comparison of the Raman spectra of sample-A1 with that from A3b and A4b (Figure 4) shows the absence of molecular hydrogen in the sample with $SiO_2$ core.

To unequivocally establish the existence of molecular hydrogen in the nanocontainer, nuclear magnetic resonance ($^1$H solid state static NMR) study was carried out to augment the Raman experiments. The peak with a chemical shift of 2.29 ppm from sample-A3b (inset to Figure 4) corroborates well with the signature of molecular hydrogen[24], thus firmly establishing our confidence in the presence of molecular hydrogen. The NMR signature from hydrogenated bulk Pd (broad peak) is also included for reference in the inset. Details related to NMR spectroscopy are considered in supplemental material (Section 2.3).

In sample-A1 (with $SiO_2$ inside the Pd shell) the hydrogen is absorbed in the Pd shell only, as the silica core is expected to accommodate negligible amounts of hydrogen; whilst in non-porous PdHS hydrogen can be stored both in the shell (atomic form) and in the interior of the Pd shell (container) in molecular form. Thus, combining the PCI and Raman data, we can now conclude that the additional hydrogen absorbed (0.11 wt.%) in sample-A4b is in the form of molecular hydrogen trapped inside the Pd shell. This leads to an enhancement in hydrogen storage capacity of 36% over nanoparticles of the same mass, along with a decrease in plateau pressure. As noted before, the Raman spectra show that molecular hydrogen is in both free and adsorbed forms in sample A3b.

The difference in the mechanism of entrapping molecular hydrogen between samples-A3 & A4 (leading to samples-A3b & A4b) is to be noted. In sample-A3b hydrogen is filled in via the pores and the pores are subsequently sealed to trap the hydrogen molecules. In sample-A4b, $H_2$ first dissociates into atomic H at the surface of Pd, followed by its diffusion into the Pd crystal. Further, the hydrogen atoms recombine at the inner surface of the PdHS to form $H_2$ molecule (Figure 5). In sample A4b the molecular hydrogen is about 18% of the total hydrogen stored. Given that the absorption studies have been carried out at 25°C, a comparison with materials like carbon nanotubes and MOFs is noteworthy. In these materials low temperatures are required (77 K) for the storage of molecular hydrogen[5]. It is to be noted that some of the observations of the current experimental study corroborate well with the computational work of Valencia et al.[17]; however, their conceptualization and configurations are considerably different.

In summary, we have developed a novel methodology to store hydrogen in nearly all conceivable methods: in a container, in molecular (adsorbed and free) and atomic (α & β solid solutions) forms. More than one method is developed to store hydrogen in molecular form, with an interesting mechanism of dissociation and recombination of hydrogen molecule being operative in non-porous PdHS samples. This strategy leads to a considerable enhancement in capacity as compared to nanoparticles of the same mass (36% increase at 1 bar, 25°C). Further, this is the first instance of storage of free molecular hydrogen in metallic nanocontainers. Molecular hydrogen is about 18% of the total hydrogen stored and this is achieved at room temperature (25°C) and 1 atm pressure. This is in stark contrast to materials like carbon nanotubes and MOFs wherein low temperatures (~77K) or high pressures are required for the storage of molecular hydrogen. Further work can take this proof of concept ahead to practical materials targeting applications.

**Acknowledgments**

Special thanks are due for Shilpi Saxena (Chemical Engineering, IITK) for help with synthesis of hollow spheres. The HPC-2010 facility of IITK is acknowledged for the computational work. Puspal Mukherjee acknowledges University Grants Commission (India) for the fellowship.

**Authors Contribution**

Anandh Subramaniam designed the present study and wrote the manuscript. Suboohi Shervani performed the following tasks: (i) synthesized the samples under supervision Sri Sivakumar and Gargi Mishra, (ii) performed TEM and Raman spectroscopy. Anshul Gupta carried out the XRD and PCI studies. Pratik Sen and Puspal Mukherjee computationally determined the Raman spectrum for various models. NMR studies were performed by Ajith Kumar and Kavya Illath. Katesh Balani worked on various aspects related to the manuscript preparation.

**FIGURES**

a   Sample A1

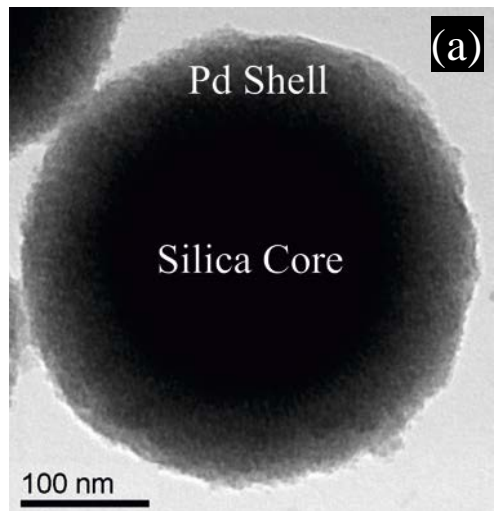

b

Sample A3

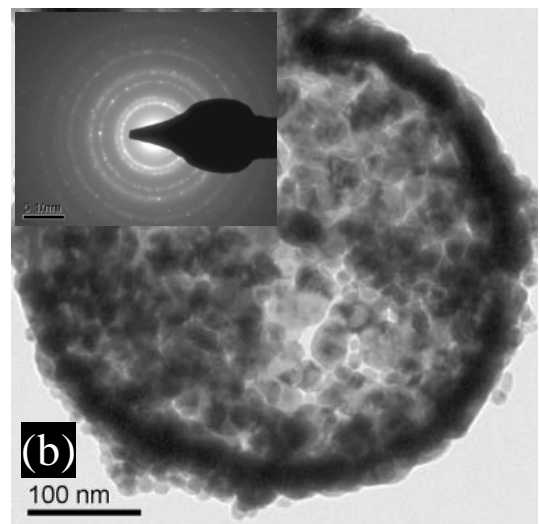

Figure 1. TEM micrographs of: (a) SiO$_2$@Pd (sample-A1) and (b) Pd hollow sphere (PdHS, sample-A3). Inset to (b) is the selected area diffraction pattern obtained from PdHS.

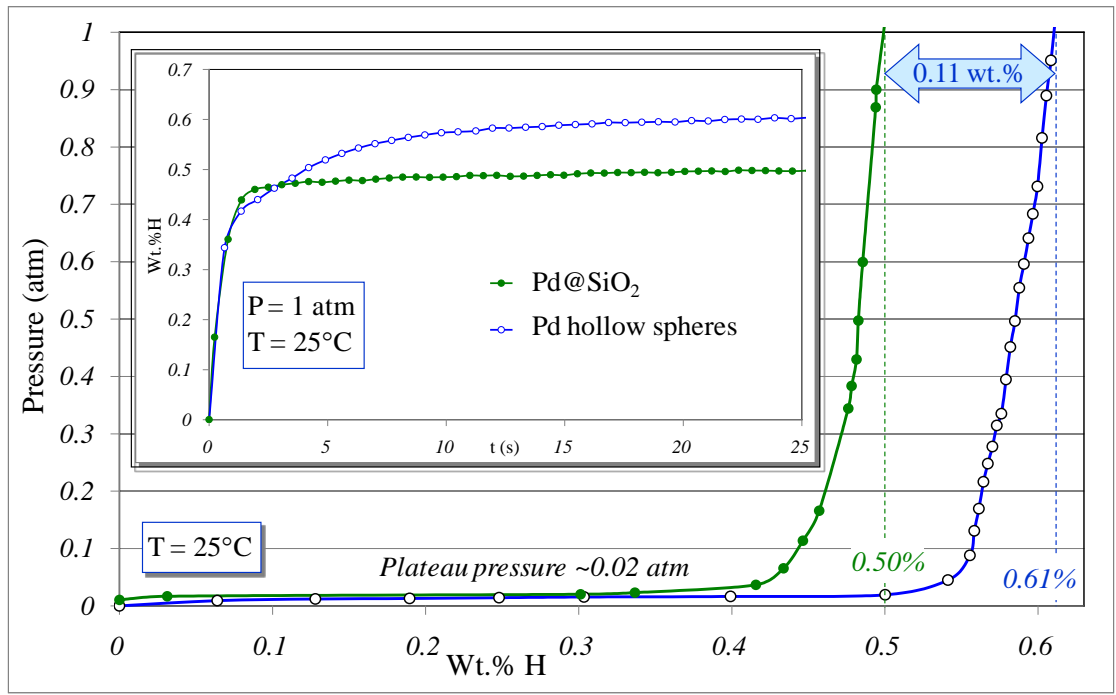

Figure 2. PCI curves obtained from: (a) Pd filled with SiO$_2$ core (sample-A1) and (b) non-porous Pd hollow sphere samples. Inset shows kinetic measurements (wt.% hydrogen absorbed with time).

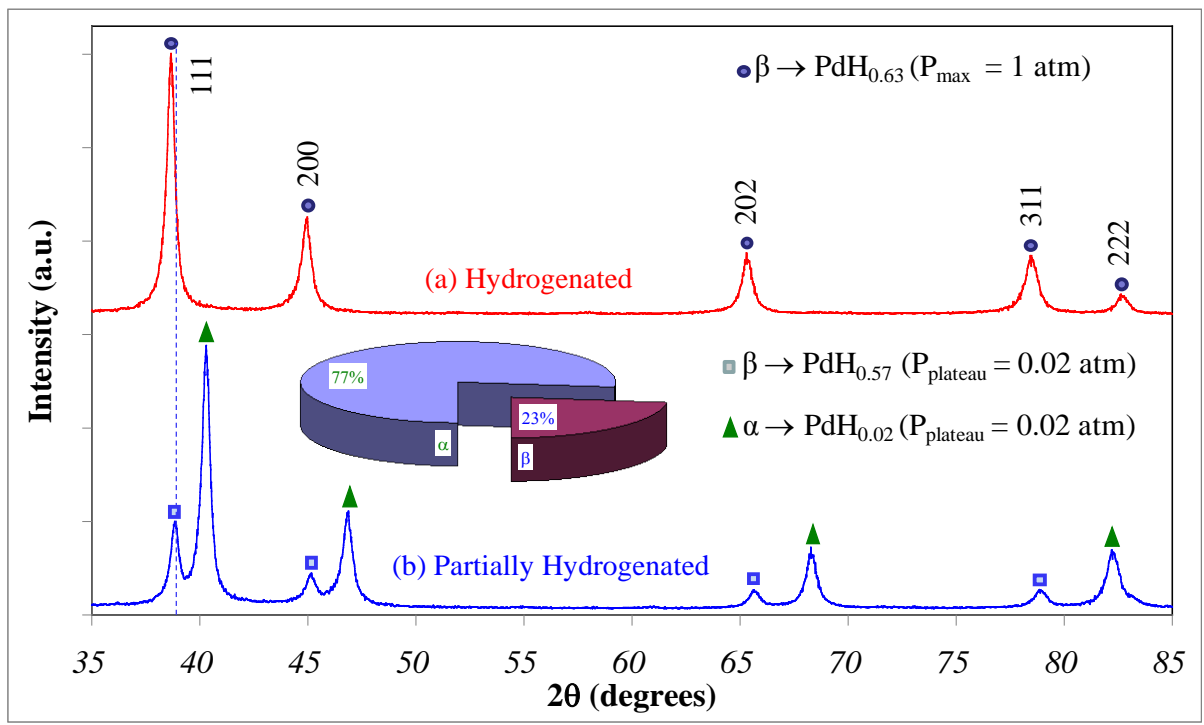

Figure 3. XRD patterns from PdHS samples (sample-A4) hydrogenated to a pressure of: (a) 1 atm, (b) 0.02 atm. The stoichiometry of the β-phase obtained in the two cases are: $PdH_{0.63}$ & $PdH_{0.57}$ respectively. Inset to (b) shows the phase fractions of α and β phases.

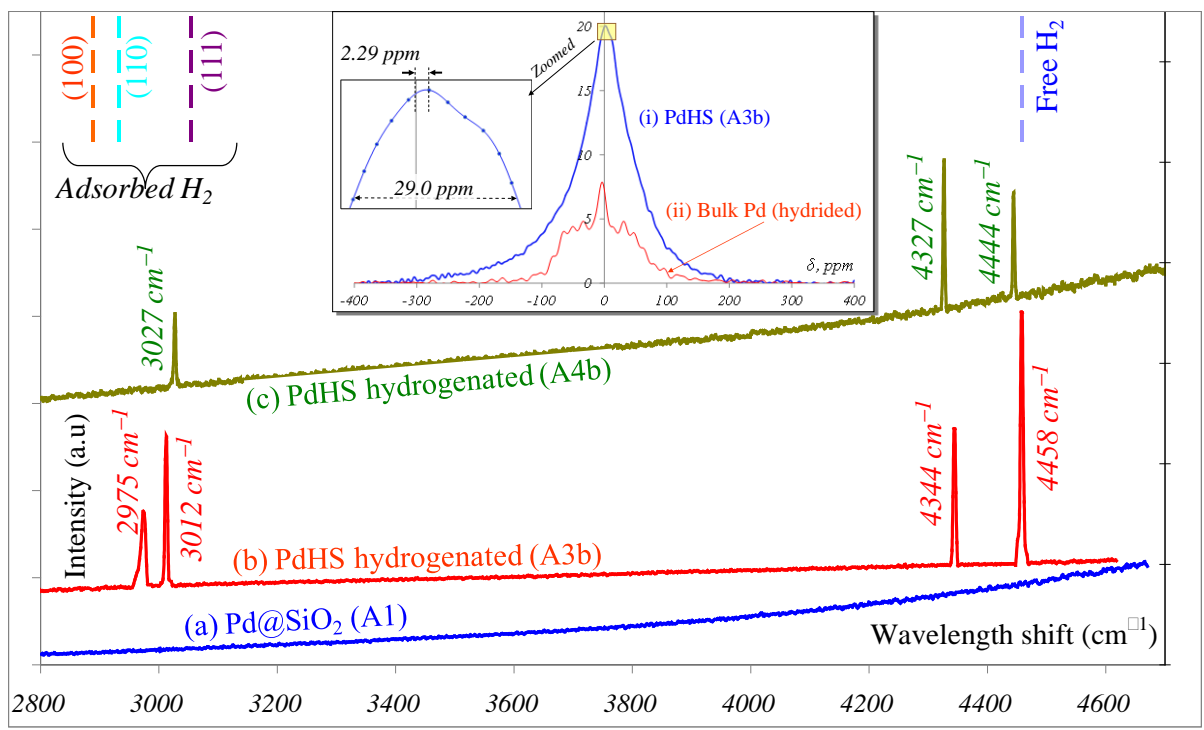

Figure 4. Raman spectra obtained from the following samples: (a) Pd filled with $SiO_2$ core (A1), (b) A3b, (c) A4b. Peak positions computed using Gaussian software is overlaid on the plots (dashed vertical lines at the top). Complete computed spectra can be found in the supplementary information. Inset shows $^1H$ solid state static NMR spectra obtained from: (i) sample-A3b and (ii) hydrogenated bulk Pd.

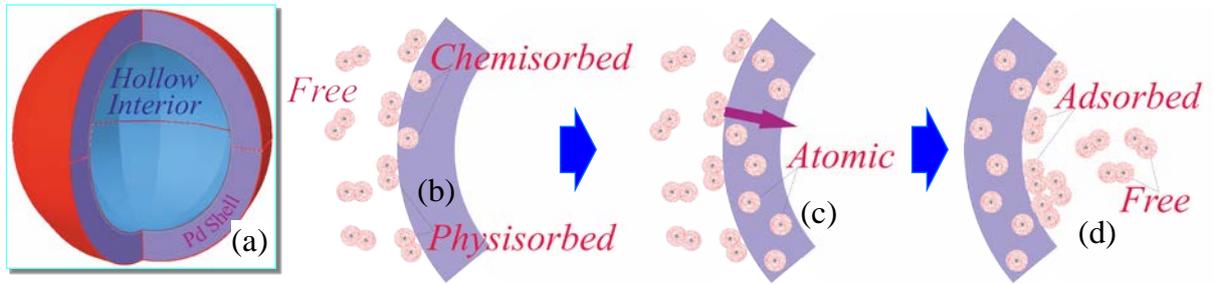

Figure 5. Schematic showing the mechanism by which hydrogen molecules are trapped inside PdHS sample-A4b (a). Steps involved are: (b) physisorption & chemisorption, (c) diffusion, (d) recombination and 'de-adsorption'.

# Multiphase Hydrogen Storage in Nanocontainers

*Suboohi Shervani[1], Puspal Mukherjee[2], Anshul Gupta[1], Gargi Mishra[3], Kavya Illath[4], Ajith Kumar[4], Sri Sivakumar[3], Pratik Sen[2], Kantesh Balani[1] and Anandh Subramaniam[1*]*

[1]Department of Materials Science and Engineering, Indian Institute of Technology, Kanpur-208016, India.

[2]Department of Chemistry, Indian Institute of Technology, Kanpur- 208016, India.

[3] Department of Chemical Engineering, Indian Institute of Technology, Kanpur -208016, India.

[4]Central NMR Facility, CSIR-National Chemical Laboratory, Pune-411 008, India

[*]anandh@iitk.ac.in *(author for correspondence)*

The supplementary materials covers the following aspects related to the manuscript on "Multiphase Hydrogen Storage in Nanocontainers": (i) finer details of the experimental methodology (Section-1) and (ii) additional results which may be of interest to a perspicacious reader (Section-2).

## 1. Methodology

The details of the experimental methods used in the main manuscript are described in this section.

### 1.1. Synthesis of Pd Hollow Nanospheres (PdHS)

Palladium hollow nanospheres were prepared using silica nanospheres as a template. Solid core mesoporous silica nanospheres (SCMS) were synthesized by modified Stober's method[1]. Palladium acetyl acetonate (100 mg) was dissolved in ethylene glycol (20 ml) and the solution was added to SCMS (100 mg) in a 1 ml plastic tube. The solution was stirred at room temperature for 2 hrs and subsequently, the mixture was heated at 230°C for 1 hr. This mixture was allowed to cool and then centrifuged at 6000 rpm for 3 min. The solid part recovered after centrifuging consists of Pd shell on silica core, which was washed in distilled water and dried (this sample is labeled A1). The silica core was removed by etching it with 2 M of NaOH for 12 h at room temperature. PdHS was recovered from the mixture by centrifugation at 7000 rpm for 3 min (sample A3).

## 1.2. Hydrogenation

Hydrogen was incorporated in two kinds of samples: porous PdHS and non-porous PdHS. Hydrogen gas of 99.99 % purity was used in the hydrogenation experiments.

### 1.2.1. Hydrogenation of Samples for Raman Spectroscopy

Hydrogenation of Sample A3 (porous PdHS) was carried out as follows. Sample was kept under rough vacuum ($10^{-5}$ bar) for 12 hours and subsequently under high vacuum ($10^{-8}$ bar) for 6 hours at 80°C (to remove all gases and volatile impurities). This was followed by pressurizing the sample under 20 bar hydrogen for 6 hours at 80°C to fill the nanospheres with hydrogen. This sample was subsequently annealed at 200°C for 2 hours to seal the pores and entrap hydrogen within the nanocontainers. The sample thus obtained is designated A3b.

Hydrogenation of non-porous PdHS was carried out as follows. As for sample A3, the sample was kept under rough vacuum ($10^{-5}$ bar) for 12 hours at 80 °C and subsequently under high vacuum ($10^{-8}$ bar) for 6 hours at 80 °C. This was followed by vacuum annealing at $10^{-8}$ bar pressure at 200 °C for 2 hours. This leads to the sealing of the pores (to obtain sample A4). The sample was subsequently pressurized at 20 bar (in hydrogen) for 6 hours at 200 °C. The sample thus obtained is designated as A4b.

## 1.3. Pressure-composition isotherms (PCI)

PCI curves were obtained from two kinds of samples: (i) Pd shell with silica core (designated sample A1) and (ii) non-porous PdHS (sample A4). Sample A1 is used as a reference sample to note the amount of hydrogen stored inside the PdHS nanocontainers.

The samples were characterized for hydrogen absorption using a commercial Sieverts apparatus (AMC Corporation, USA) to obtain pressure–composition isotherms (PCI) and to acquire kinetics data (wt.%H absorbed vs. time plots). Details of the experimental apparatus have been given elsewhere[2]. Temperature of the system was measured using type K thermocouple in conjunction with Omega Controllers (CN8241) with an accuracy of ±0.2 °C under equilibrium conditions and stability better than ±0.05 °C. Pressure measurements were carried out using pressure transducers (Honeywell TJE) which have resolution of 0.5 mbar. Approach to equilibrium was monitored using the change in slope of temperature with time (0.002 °C/s) and pressure with time (0.00245 psi/s) in the sample chamber. The equilibrium of the gas is considered fully established when these two values together reach nearly zero for at least 2 min.

Pressure-composition isotherm measurements were carried out at 25 °C from $10^{-3}$ to 1 bar. Kinetics of hydrogen absorption was measured at 25 °C and 1 bar initial pressure. Amount of hydrogen absorbed was measured per unit weight of palladium for both PdHS (sample A4) and Pd filled with Silica (sample A1).

### 1.4. X-ray diffraction

X-Ray diffraction measurements were made using PANalytical Empyrean equipment with 2θ ranging from 35° to 130° at a step size of 0.015° (200 s per step) using Cu Kα radiation for the PdHS sample. The scan range was limited to 85° (400 s per step) for the hydrogenated samples for better accuracy. The x-ray source was operated at 45 kV and 40 mA voltage and current ratings respectively.

Lattice parameters and phase fractions were determined using Whole Profile Fitting (also known as Pawley method, the basis of the Rietveld method) feature of JADE software supplied by Materials Data Incorporated (MDI)[3]. Structureless refinement has been carried out using reference patterns from PDF 2 database (complete with d's and I's values along with lattice constants) without refining crystal structure (atomic coordinates etc.).

### 1.5. Transmission electron microscopy (TEM)

TEM studies were performed using High Resolution Transmission Electron Microscope (HR-TEM, FEI Titan G2 60−300 TEM) instrument operated at 300 kV. Samples A1 & Sample A3 were prepared using 10 mg of sample dispersed in 2 ml ethanol using ultrasonication for 30 minutes. The above mentioned samples were deposited on carbon coated Cu grids of 300 mesh size (Ted Pella, Inc. USA). The particle size and shape were determined using bright field images (BFI), while selected area diffraction patterns (SADP) were obtained to investigate the crystalline nature of the sample.

### 1.6. Raman Spectroscopy

Raman spectroscopy was performed using high-resolution confocal micro Raman spectrometer (STR-300) with spectral resolution less than 0.5 cm$^{-1}$ and equipped with a cooled CCD camera. The focal length of the spectrometer is 300 mm and He-Ne LASER of wavelength of 633 nm was used in the studies. Data was obtained in the range 3000-5000 cm$^{-1}$ (using a grating having groove density of 600 per mm). The temperature of the ambience (room temperature) was maintained at 18°C during the measurements. About 500 mg of sample was spread across a 2 mm cross sectional area of a glass slide and this was exposed to the LASER radiation for 30 s in each accumulation. A total of 30 such data accumulations were obtained per sample.

**1.6.1. Computation of Raman spectra**

To identify the origin of the peaks in the experimental Raman spectrum, computation of Raman spectra was performed using Gaussian-09 software [4]. Three kinds of configurations have been studied: (i) free molecular hydrogen, (ii) hydrogen adsorbed on crystalline surfaces (physisorbed hydrogen on metal and metal hydride substrates) and (iii) weakly adsorbed hydrogen (as in second and higher layers in multi-layer adsorbates). In point (ii) above, the numbers of configurations possible are extremely large, which is based on the crystallography and defect structure of the surface. These complications include: (a) the hydrogen could be adsorbed on (111), (110), (100) or other surfaces, (b) there could be ledges and kinks on these surfaces, (c) the substrate surface could have chemisorbed hydrogen atoms of varying stoichiometry. However, given the reasonably sharp peaks observed in the experimental Raman spectra, it seems that some of these configurations are dominant over the others. Keeping this in view, selected (model) configurations of adsorbed hydrogen on substrates have been used in the computation of the Raman Spectra. These are: (i) (111) surface of Pd, (ii) (110) surface of Pd, (iii) (100) surface of Pd, (iv) (111) surface of $PdH_{0.54}$. In addition spectra were computed for pure Pd and $PdH_{0.54}$.

Computations for hydrogen molecule adsorbed on metal surface (having (111), (110) or (100) orientation) was performed using ~20 Pd atom metal clusters (keeping at least six Pd atoms on the top layer, with at least three layers of atoms). As a first step the geometry of the Pd clusters was optimized using density functional theory (DFT), utilizing the hybrid functional of Truhlar and Zhao M06 [5]. The Los Alamos ECP plus DZ (LANL2DZ) basis set was used for Pd atoms and 6-311G++(d,p) for hydrogen atoms. Pseudo keyword was used to substitute a model potential for the core electrons. In addition, tight optimization criteria and ultrafine pruned grid was used for computing very low frequency modes. This is because the system contains numerous tetrahedral centers. Once the structure has been optimized, the same conditions (the basis set and functional) were used to compute the frequency modes of adsorbed hydrogen molecule. The model was used for the computation of both Raman and IR vibrational spectra, however in this work we have reported only the Raman frequencies. The final orientation of the $H_2$ molecule on the Pd surface was arrived at by starting the same calculations from three initial geometries of the $H_2$ molecule: (i) perpendicular, (ii) parallel and (iii) random orientation. It is observed that all the initial geometries led to almost same final position of the molecule and the energy computed for the three cases are the same. Once the structure is optimized, the frequencies are computed and the normal mode of $H_2$ stretching is visualized using the GaussView software [6]. This protocol (as outlined above) is followed for the computation of spectra for: (a) palladium hydride crystal and (b) $H_2$ molecule adsorbed on palladium hydride crystal.

### 1.7. Nuclear magnetic resonance

Solid- State $^1$H-NMR studies were performed using Bruker NMR spectrometer (300 MHz, 7.05 T). $^1$H-NMR shifts were measured in parts per million (ppm), downfield from the trimethylsilane (TMS) standard sample. NMR experiments were performed on sample A3b (synthesis as described in 1.2.1) at 303 K and 328 K. The samples were placed in 3.2 mm OD and 15.4 mm long Zirconia rotors. In addition, hydrogenated bulk palladium sample was used as reference for comparison with the A3b sample. In this sample, unlike the A3B sample, there is no void space for hydrogen molecules to accumulate. It is to be noted that all samples were handled in glove box (after synthesis and during transfer) to prevent the hydrogen desorption, contamination and oxidation.

## 2. Results and Discussions

### 2.1. X-Ray diffraction

#### 2.1.1. PdHS

Figure S1 shows X-Ray diffraction pattern from PdHS (sample A3). The peaks have been indexed to single phase palladium. Crystallite size and strain, obtained from Williamson Hall Plot (shown in inset), are 29.8 nm and 0.186%, respectively.

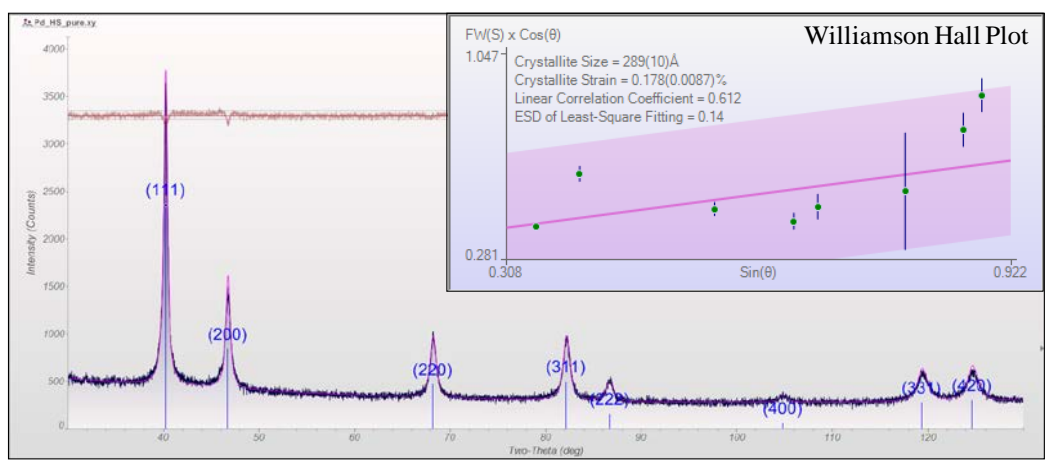

Figure S1 X-Ray diffraction pattern from PdHS (sample A3). Inset shows Williamson Hall Plot used to calculate crystallite size and strain.

### 2.1.2. Pd@SiO$_2$

Figure S2 shows X-Ray diffraction data from Pd@SiO$_2$ (sample A2). Blank scan of amorphous (glass) sample holder was used as background to distinguish from the amorphous signature from the sample. Pseudo-Voigt profile was used to fit the amorphous hump (profile fitting parameters shown in inset) using WPF in JADE. Density of Pd (crystalline) and SiO$_2$ (amorphous) are assumed to be equal to literature values (12 g/cc and 2.6 g/cc, respectively) in the phase fraction calculations[7].

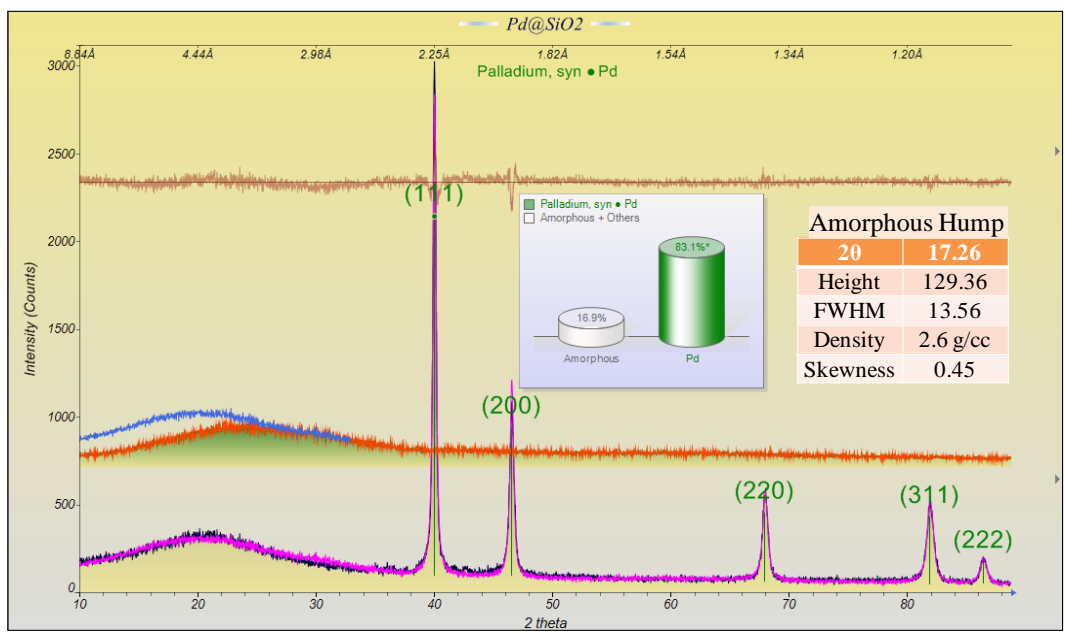

Figure S2 X-Ray diffraction data obtained from Pd@SiO$_2$ (sample A2) with phase fraction calculated using WPF (JADE, MDI Corp). Also shown is the deconvolution of the background profile from the amorphous profile of the sample (in orange and blue color respectively). Schematic on the right is used to theoretically calculate the phase amount.

### 2.1.3. PdHS (hydrogenated)

The results of Rietveld refinement using Whole Profile Fitting (WPF) in JADE is shown in Figure S3. The pink color plot represents fitted profile overlaid on the experimental x-ray pattern. The plots in green and blue represent the individual profiles of palladium (α) and palladium hydride (β) respectively, deconvoluted from the experimental pattern. The details for reference patterns are listed in Table S1.

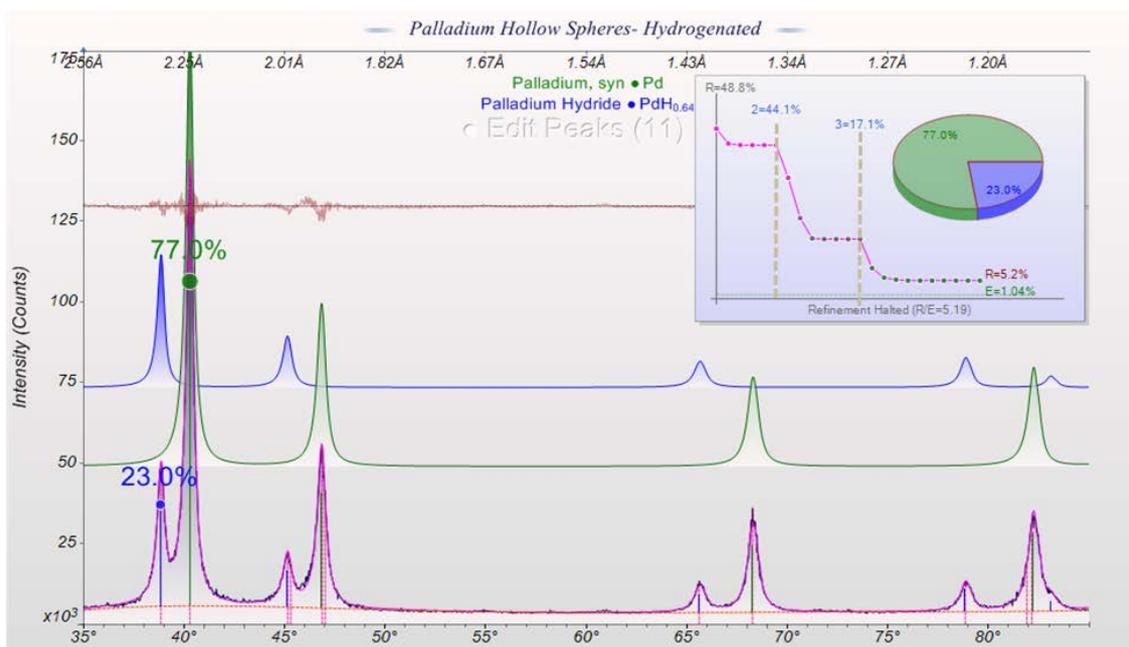

Figure S3 Rietveld refinement of XRD data obtained from hydrogenated palladium hollow spheres (sample A3b) sample using Whole Profile Fitting (WPF) feature of JADE. The plot of R% and phase fraction (pie chart) is shown in the inset.

In the refinement shown in Figure S3, a fifth order polynomial is used for fitting the background curve. The dashed grey vertical markers in the R% plot (inset to the Figure S3) indicate the beginning of each refinement round. The results of the Refinement for Profile Shape Function are given in Table S2 and the corresponding phase analysis is given in Table S3.

Table S1 Structural data used for Whole Profile Fitting in JADE.

| Phase | Chemical Formula | PDF# | Space Group | RIR |
|---|---|---|---|---|
| Palladium | Pd | 00-046-1043 | cFm3m (225) | 17.00 |
| Palladium hydride | $PdH_{0.64}$ | 01-084-0300 | cF23 (196) | 16.21 |

Table S2 Parameters obtained after refinement for profile shape function using whole profile fitting in JADE.

|  | Skewness |  | Lorentzian Component |
|---|---|---|---|
| s0 | 1.47343 | p0 | 0.68456 |
| s1 | -2.68324 | p1 | 0.63414 |
| s2 | 1.32353 |  |  |

Table S3 Phase analysis using Whole Profile Fitting in JADE.

| Sample | Phase | Bragg R% | Phase fraction (wt.%)# | Lattice Parameter (Å)# | Chemical formula | Composition* at.%H | Composition* wt.%H |
|---|---|---|---|---|---|---|---|
| PdHS | Palladium | 6.80 | 100 (1.9) | 3.890 (12) | Pd | - | - |
| PdHS (partial hydrogenated) | Palladium (α) | 7.50 | 77 (4.5) | 3.893 (14) | $PdH_{0.02}$ | 1.96 | 0.018 |
|  | Palladium hydride (β) | 9.87 | 23 (8.7) | 4.023 (15) | $PdH_{0.57}$ | 36.30 | 0.53 |
| PdHS (hydrogenated) | Palladium hydride (β) | 7.98 | 100 (1.6) | 4.032 (13) | $PdH_{0.63}$ | 38.65 | 0.59 |

# *The figures in the bracket indicate Estimated Standard Deviation (ESD).*

*Composition of the solid solution and hydride phase in hydrogenated palladium hollow spheres was determined using a linear relation between lattice parameter and hydrogen concentration as derived from Baranowski et al [8]:*

$$a^3 = 10.5x + a_0^3$$

*where $a_0$ represents lattice parameter of palladium hollow spheres upon hydrogen absorption (i.e. as compared to the non-hydrogenated sample) and x represents fraction of hydrogen with respect to Pd during hydrogen absorption.*

**2.2. Computation of Raman Spectra**

Figure S4 shows Raman spectra computed for various configurations. The vibrational peaks observed in these spectra are tabulated in Table S4. A comparison with the experimental Raman spectra (Fig. 4) indicates the presence of free and adsorbed molecular hydrogen. It is also clear that there is no signature from: (i) $H_2$ adsorbed on (111) surface of $PdH_{0.54}$, (ii) from crystalline Pd or (iii) crystalline $PdH_{0.54}$.

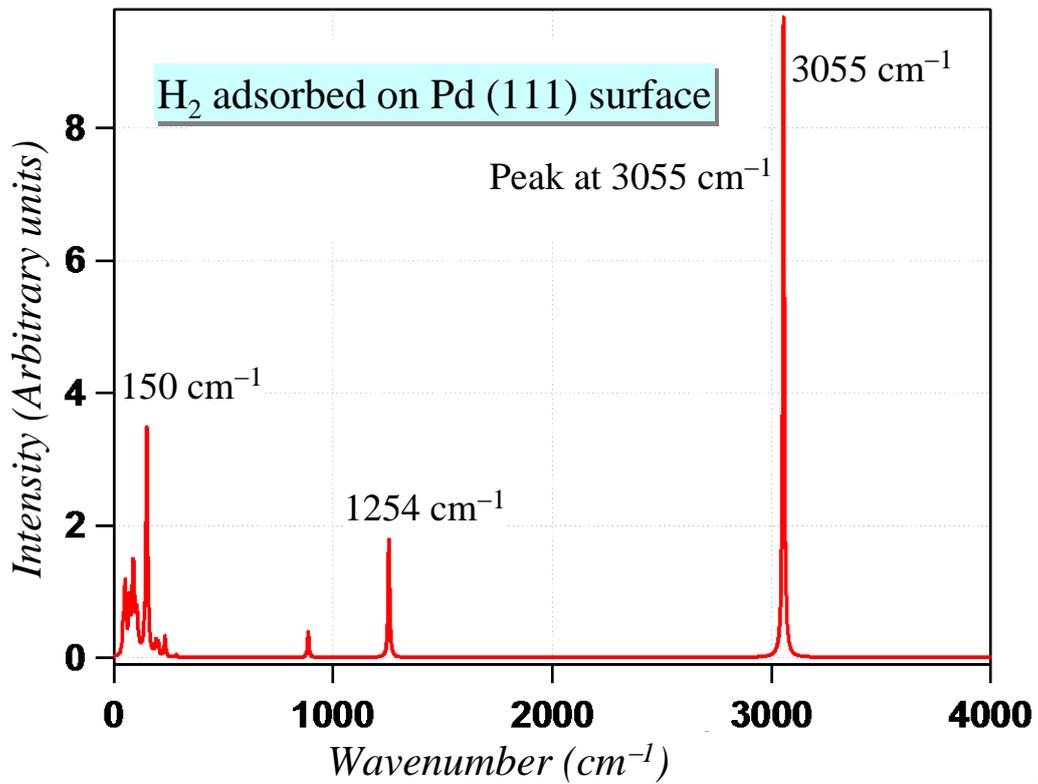

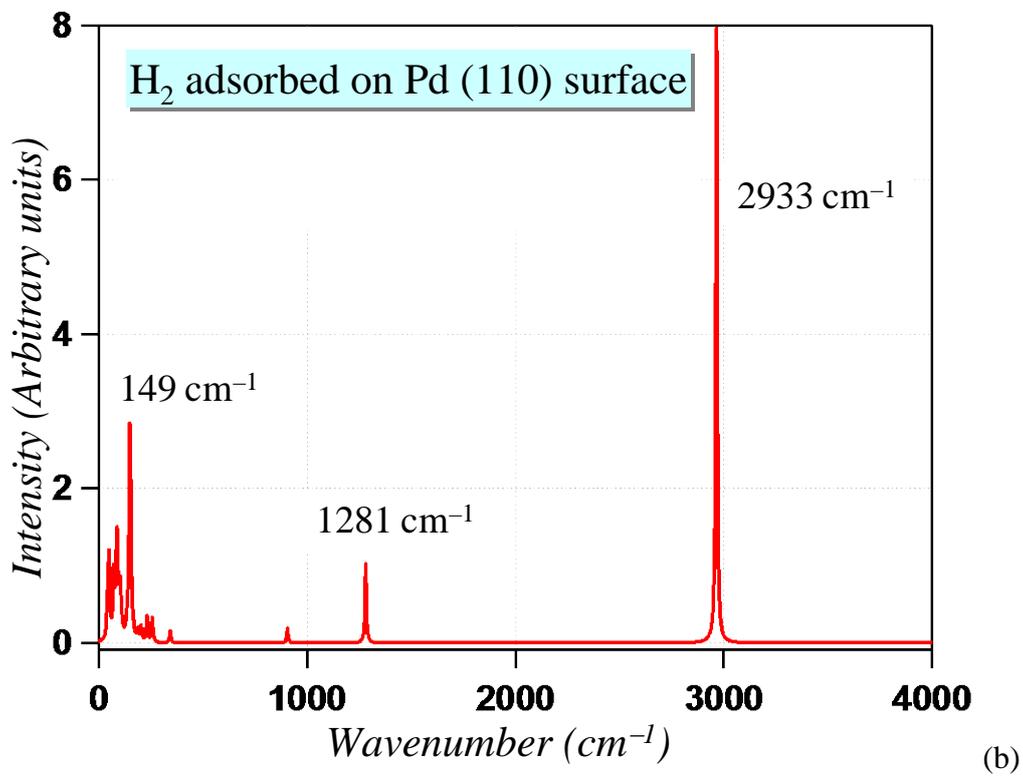

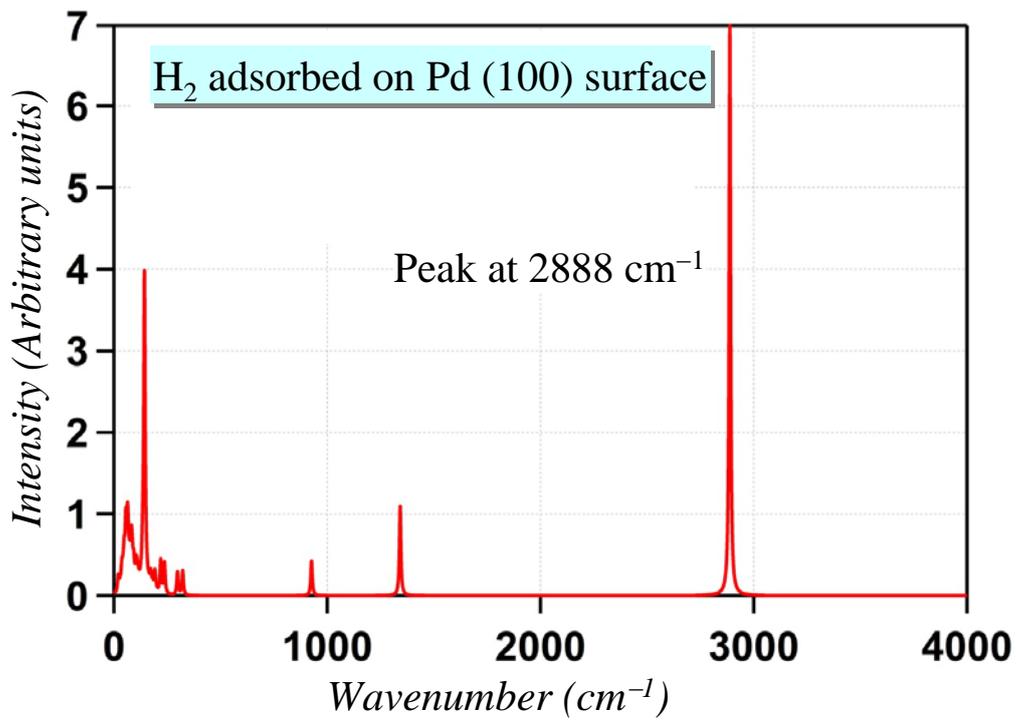

(c)

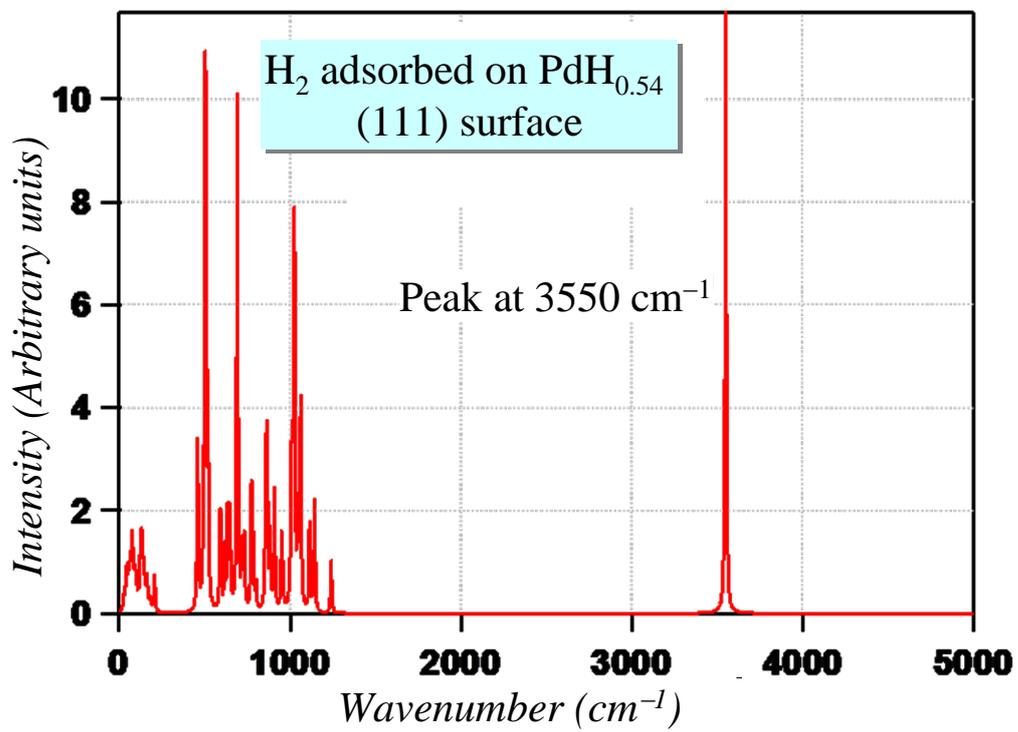

(d)

(e)

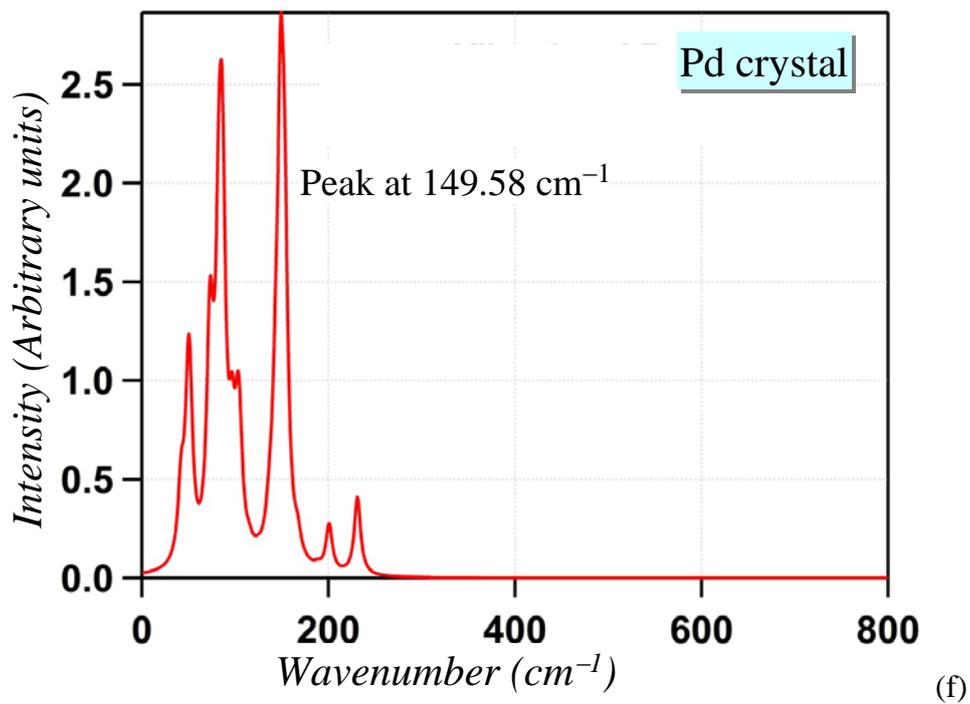

(f)

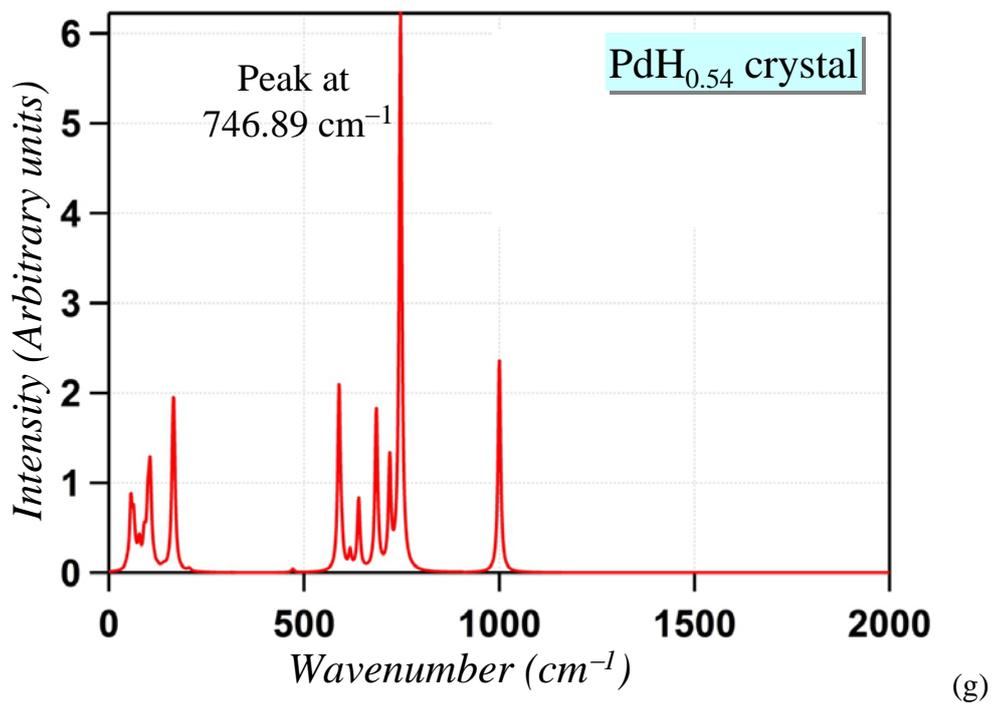

(g)

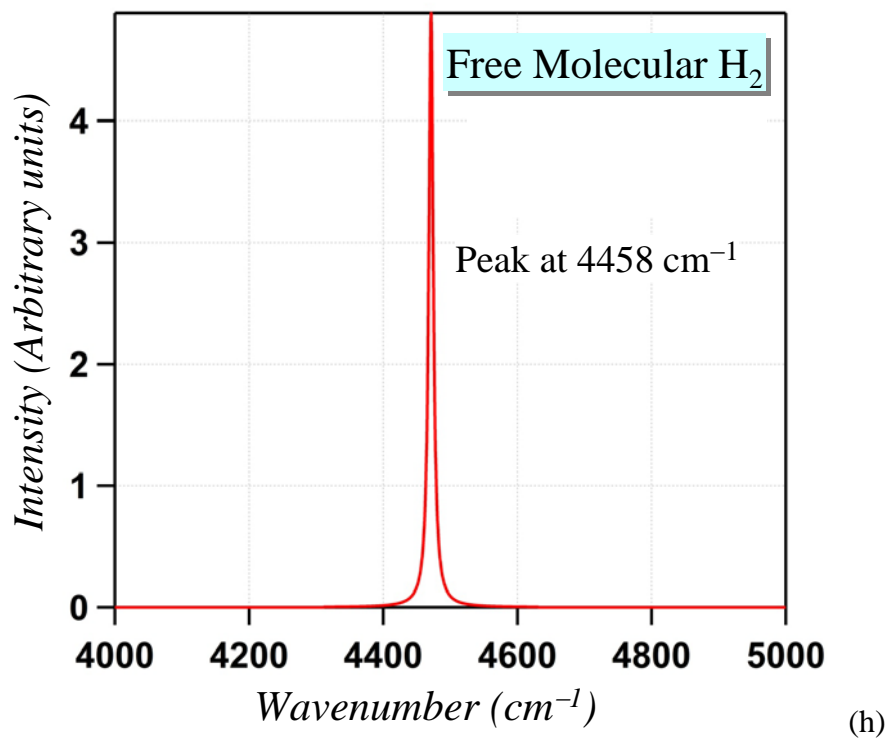

(h)

Figure S4 Raman Spectra computed using Gaussian software for: (a) $H_2$ adsorbed on (111) surface, (b) $H_2$ adsorbed on (110) surface, (c) $H_2$ adsorbed on (100) surface, (d) $H_2$ adsorbed on (111) surface of $PdH_{0.54}$, (e) weakly adsorbed $H_2$ (as in a multilayer adsorbate), (f) Pd crystal, (g) $PdH_{0.54}$ crystal, (h) Free Molecular $H_2$ (as inside the nano-container).

Table S4  Wavelength shift observed in the computed Raman Spectra.

| Configuration | Details | Vibrational peaks *Wavenumber (cm$^{-1}$)* |
|---|---|---|
| Free $H_2$ | - | 4458 |
| Pd | bulk | 149.58 |
| $PdH_{0.54}$ | bulk | 746.89 |
| $H_2$ adsorbed on Pd | (111) surface | 3055 |
| | (110) surface | 2933 |
| | (100) surface | 2888 |
| $H_2$ adsorbed on $PdH_{0.54}$ | (111) surface | 3550 |
| Weakly adsorbed $H_2$ | As in a multilayer adsorbate | 4329 |

## 2.3. NMR Spectroscopy

Solid state $^1H$ NMR experiments performed at 328 K show chemical shift at 4.49 ppm (FWHM 66.0 ppm), as compared to chemical shift at 2.29 ppm (FWHM 92.0 ppm) for room temperature measurements (303 K) (Figure S5). The measurements are relative to the standard TMS sample. It is seen that on heating chemical shift increases, while the FWHM decreases. These results not only indicate the presence of molecular hydrogen in both samples[9], but also suggest that on heating, a fraction of the hydrogen molecules adsorbed on the inner surface of palladium hollow sphere become free within the sphere.

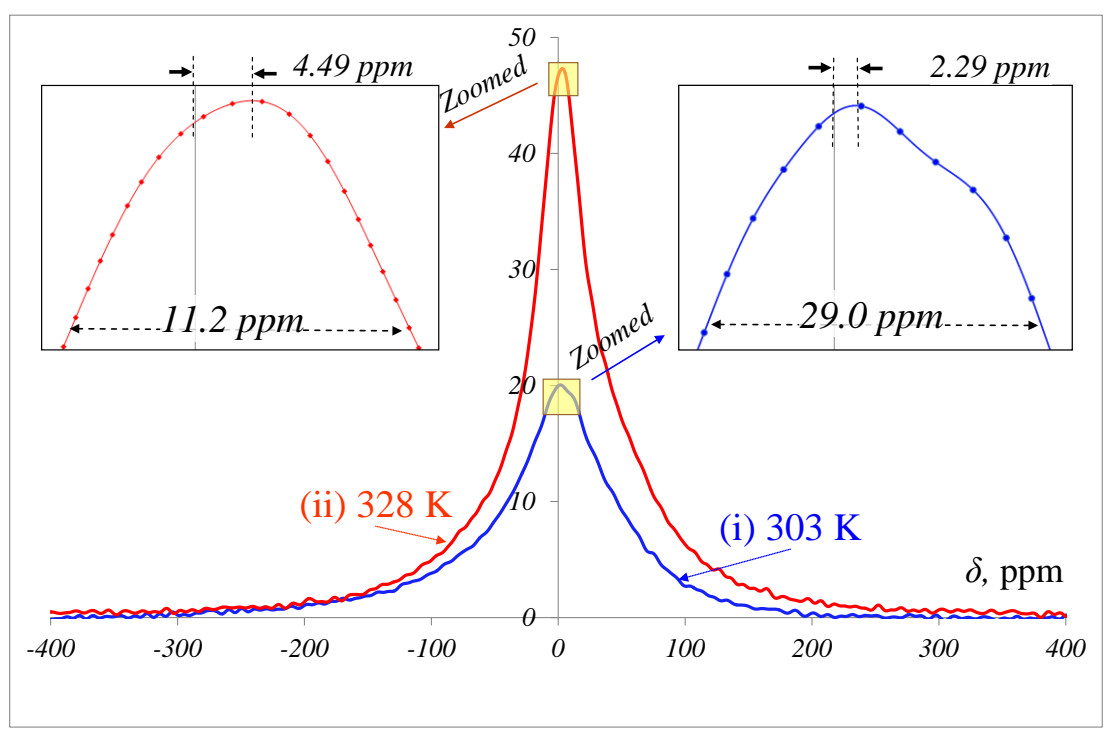

Figure S5 Solid state $^1H$ NMR spectra of sample A3b at: (a) 303 K & (b) 328 K.